\documentclass{svproc}

\usepackage{url}
\usepackage{graphicx}
\usepackage{multicol}
\usepackage[bottom]{footmisc}
\usepackage{tikz}
\usetikzlibrary{calc}
\usepackage{float}
\usepackage{amsmath}
\usepackage{subcaption}
\usepackage{amssymb}
\usepackage{amsfonts}
\usepackage{svg}
\usepackage{cite}
\usepackage{hyperref}

\usepackage[font=footnotesize]{caption}

\graphicspath{{./Figures}}

\begin{document}
\mainmatter

\title{The Effect of Gait Stability Based on Two Types of Impact Strategies for Two-Link Walking and Brachiating Robots}

\titlerunning{Gait Stability Based on Impact Strategy}  

\author{Alan Estrada Flores\inst{1} \and Nelson Rosa Jr.\inst{1}}

\authorrunning{Alan Estrada Flores \and Nelson Rosa Jr.}

\tocauthor{Alan Estrada Flores and Nelson Rosa Jr.}

\institute{Illinois Institute of Technology, Chicago IL 60616, USA}

\maketitle

\section{Introduction}
\begin{figure}[t]
\centering
\includegraphics[width=0.7\linewidth]{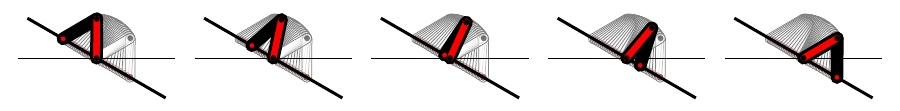}
\hfill
\includegraphics[width=0.7\linewidth]{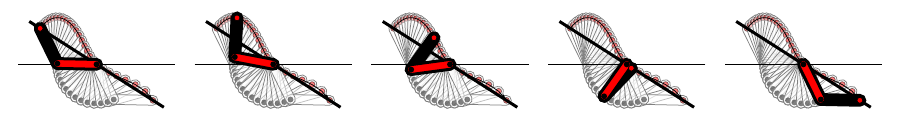}
\hfill
\includegraphics[width=0.7\linewidth]{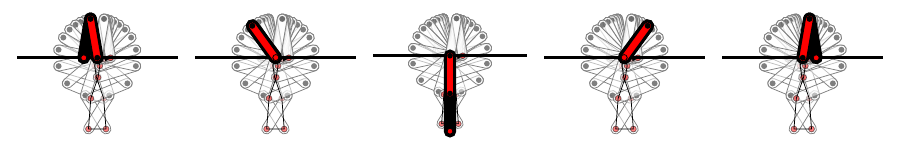}
\hfill
\includegraphics[width=0.7\linewidth]{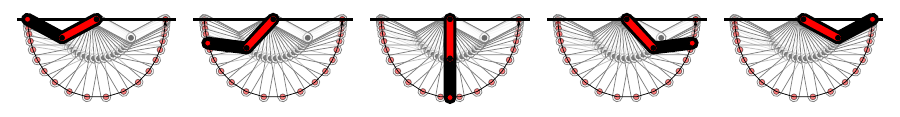}
\caption{Example period-one walking and brachiating gaits of a two-link robot.}
\label{fig:motions}
\end{figure}

The two-link walking models of \cite{Garcia1998} and \cite{Goswami1998} and the
brachiating model of \cite{Gomes2005} have provided valuable insight into the
mechanics of limbed locomotion in the fields of biomechanics and robotics.
Figure~\ref{fig:motions} shows example motions of the model in \cite{Gomes2005}.
In addition to contributions to human and robot gait analysis, these works show
that the underlying impulsive dynamics exhibit rich nonlinear phenomena.

For bipedal walking, these results include the existence of passive dynamic
walking gaits (i.e., unactuated periodic walking motions of the impulsive
dynamics), period-doubling bifurcations, strange attractors, and chaotic motion
\cite{Garcia1998, Goswami1998}.  A large body of work has since explored
variations of these canonical walking models with the addition of, e.g., springs
and dampers, limbs, and forcing functions (see \cite{Safartoobi2022} and
references therein).  Common results include the existence of period-doubling (PD)
bifurcations and their associated route to chaos.  The results for brachiation
are much more sparse, but a core result is the existence of unstable passive
dynamic gaits with zero energy loss across impacts and
energetically-conservative swing dynamics \cite{Gomes2005}.  These results have
been extended to show the existence of stable and unstable passive dynamic and
actuated gaits and observations of fold (FD) and Neimark-Sacker (NS) bifurcations
\cite{Rosa2012, Rosa2013b}.

Given past work, a gap that still persists is the lack of a unified bifurcation
analysis of two-link models and their set of walking and brachiating gaits
\cite{Rosa2013b}.  Ultimately, both of these fields study the same equations of
motion, but results are only reported in isolation.  Specifically, both fields
analyze the behavior of fixed points of a discrete step-to-step map.  The map
represents the continuous swing motion of the robot as it pivots about its hand
or foot followed by an instantaneous plastic collision with an inclined surface.
Because of the bilateral symmetry of many two-link models, a periodic motion
(i.e., a gait) can be identified after just one step and a relabeling of
coordinates \cite{Garcia1998, Goswami1998, Safartoobi2022, Rosa2012, Rosa2013b}.
The periodicity of a fixed point is thus defined with respect to the number of
steps the robot takes.  In other words, a gait that takes $n$ steps is a
period-$n$ fixed point.

Another understudied facet in both fields is the switching function of the
impulsive dynamics.  In our numerical study, we explore a step-to-step map under
state-based switching (SBS) and time-based switching (TBS) strategies.  In SBS, a collision occurs whenever the distance between the surface and the
end effector of the robot (serving as a foot for walking or hand for
brachiating) is zero.  In the literature, SBS is the de facto choice with the
slope of the surface treated as a free parameter of the system.  For TBS,
collisions occur after a pre-determined duration of time has elapsed.  In this
case, switching time is the free parameter.  TBS is a viable control option for
robots in vertical environments (e.g., \cite{Rosa2012,Mbewe2025}) as contacts
can be made at any point in time on the surface.

As further motivation for this study, an interesting aspect of switching strategies is that both strategies produce the same set of period-one gaits.  We omit a
formal proof as this is intuitively true.  For every gait under SBS, we can
take a stopwatch to determine the switching time, and, for every gait under TBS,
we can record the slope at the switching time.  Given this fact, we still
observe different long-term gait behaviors across impacts depending on the
switching strategy.  This leads to our core contributions.

\subsection{Statement of Contributions}
In this paper, we explore the impulsive dynamics common to single-joint,
two-link models of walking and brachiating gaits with respect to slope and
switching time.  In particular, we expand on the observations made in
\cite{Rosa2013b} with respect to how the stability of a gait and bifurcations
encountered within a family of gaits change under TBS and SBS of the impulsive
dynamics.  Our contributions are
\begin{enumerate}
\item \textbf{A numerical study of gait stability under two switching
strategies}.  We study the stability of gait families with respect to
time-based and state-based switching functions of the impulsive dynamics.  When
there is a loss of stability in a gait family, we observe PD, NS, and FD bifurcations.

\item \textbf{An exploration of the role of symmetry with gait families}.
Two-link models with bilateral symmetry have period-one gait families that are
mirror images of each other about the switching time axis.  We investigate the
evolution of unstable gaits in these families for one such model.  In
particular, we find that near FD bifurcations, unstable gaits that start in one
gait family converge to stable gaits in the mirrored family of gaits under TBS.

\end{enumerate}

\begin{figure}[t]
\centering
  \begin{tikzpicture}
    \node[anchor=center] (basicPDB) at (-3, 0) {\includegraphics[scale=0.25]{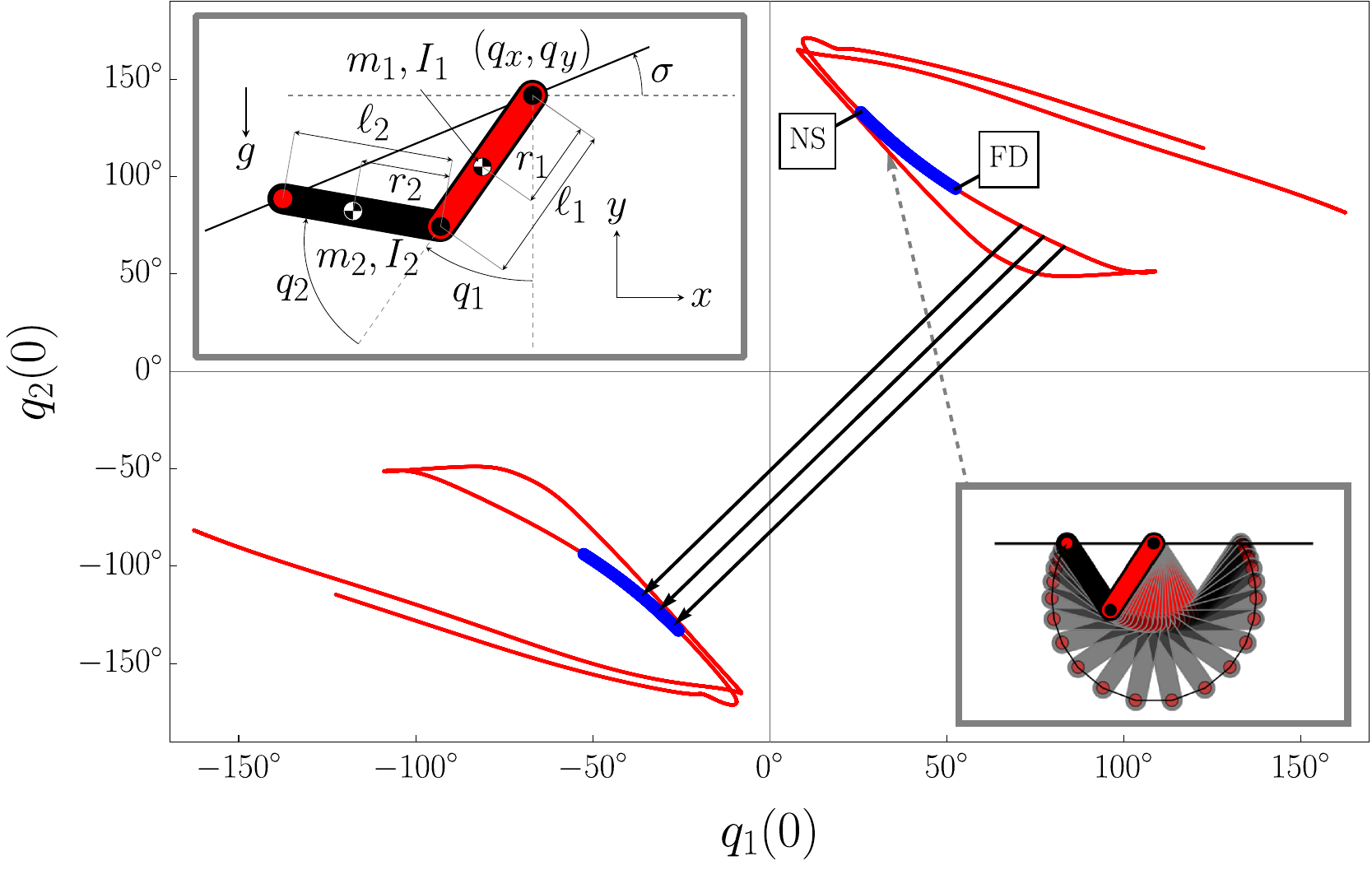}};
    \node[anchor=center] (mainBase) at (2.95,-0.02) {\includegraphics[width=0.37\textwidth]{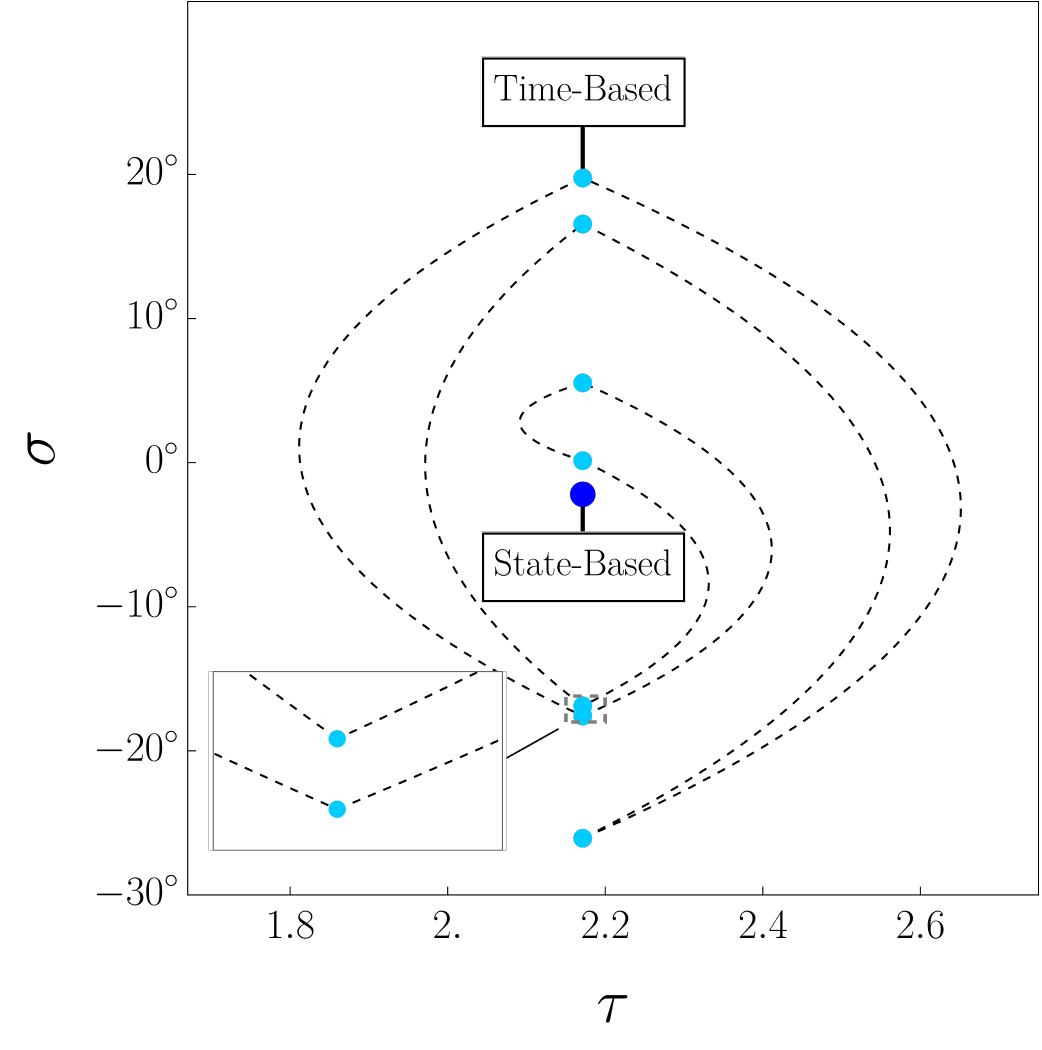}};
  \end{tikzpicture}
\caption{(left plot) Curves representing two symmetric gait families projected
onto the $q_1(0)$-$q_2(0)$ (the configuration $q(t)$ at $t = 0$) plane under a TBS strategy.  Red
represents unstable gaits and blue stable gaits.  The black arrows depict
unstable gaits on one family converging to stable gaits on the mirrored family.
Each curve has NS and FD bifurcation points.  Finally,
the top-left quadrant shows the model, and the bottom-right quadrant shows an example
brachiating gait on a curve.  Its mirrored gait (not shown) brachiates in the
opposite direction. (right plot) The evolution of the example brachiating gait,
which is unstable.  Under SBS, it converges to a stable
period-one fixed point.  With TBS, it converges to a stable
period-seven fixed point.}

\label{fig:two_link_model}
\end{figure}

Figure~\ref{fig:two_link_model} provides a visual summary of these
contributions.

\section{The Two-Link Model}
\label{sec:background}
Figure~\ref{fig:two_link_model} shows the coordinates and parameters of our
two-link model under a constant gravitational force $g$ and surface incline
$\sigma$.  The parameters $\ell_i$ and $r_i$ ($i \in \{1, 2\}$) represent the
lengths of the links and positions of the centers of mass, respectively.  The
parameters $m_i$ and $I_i$ are the masses and moments of inertia about the
center of mass of each link, respectively.  The full configuration vector is
$q{'} = [q_x, q_y, q_1, q_2]^T \in \mathbb{R}^4$, where $(q_x, q_y) \in
\mathbb{R}^2$ are the $(x, y)$ world coordinates of the model, $q_1$ is the
angle of link 1 from the vertical, and $q_2$ is the angle of link 2 with respect
to link 1.  The pivot point $(q_x, q_y)$ is fixed throughout the swing motion.
At impact, the post-impact values are fully determined by the pre-impact values
of $q_1$ and $q_2$.  For convenience, we use a reduced configuration vector $q =
[q_1, q_2]^T \in \mathbb{R}^2$ with state vector $x = [q^T, \dot{q}^T]^T \in
\mathbb{R}^4$.

In reduced coordinates, trajectories satisfy the equations of motion
$\mathcal{H}$:
\begin{equation*}
\mathcal{H} = \begin{cases}
\dot{x}(t) = f(x(t)) & \phi(x(t), t) \neq 0 \\
x^+ = h(x(t)) & \phi(x(t), t) = 0,
\end{cases}
\label{eqn:eom}
\end{equation*}
where $x^+ \in \mathbb{R}^4$ is the post-impact state, $f(x) = [\dot{q}^T,
\ddot{q}^T]^T \in \mathbb{R}^4$ is a continuous vector field, $h(x) \in
\mathbb{R}^4$ is a jump map, and $\phi(x, t) \in \mathbb{R}$ is a switching
function.  We define $f(x)$, $h(x)$, and $\phi(x, t)$ in
Sections~\ref{ssec:cd}--\ref{ssec:switch}.

When pieced together, the complete hybrid system can be written as a discrete
step-to-step map $x_1 = h(x(\tau, x_0))$, where $x_0$ and $x_1$ are post-impact
states at time $t = 0$ and $t = \tau$, respectively; $x(t, x_0)$ is the
associated flow of the continuous dynamics such that
\begin{equation*}
x(t, x_0) = x_0 + \int_0^t f(x(s))\;ds;
\end{equation*}
and $\tau \in \mathbb{R}$ is a switching time such that $\tau = \inf\{t > 0:
\phi(x(t, x_0), t) = 0\}$, i.e., $\tau$ is the first switching time of
the hybrid trajectory starting from $x_0$ at $t = 0$.

In terms of locomotion, a gait corresponds to a fixed point of the impulsive
dynamics $\mathcal{H}$.  For example, a period-one fixed point satisfies $x_1 =
h(x(\tau, x_0^*)) = x_0^*$.  More generally, a period-$k$ gait satisfies $x_k =
h(x(\tau, x_{k-1})) = x_0^*$.  We are mainly concerned with tracing period-one
fixed points and studying their stability properties.  This leads us to our
periodicity map $P(x_0, \tau) \in
\mathbb{R}^4$:
\begin{equation}
P(x_0, \tau) = h(x(t, x_0)) - x_0.
\label{eqn:P}
\end{equation}
For period-one fixed points, we trace the roots of $P(x_0, \tau) = 0$ using
numerical continuation methods \cite{Allgower2003}.

\subsection{The Continuous Dynamics}
\label{ssec:cd}
The continuous dynamics of the model during the swing phase are those of a
double pendulum.  The equations can be written in standard form for mechanical
systems and solved for $\ddot{q}$ such that
\begin{equation*}
M(q(t))\ddot{q}(t) + C(q(t), \dot{q}(t))\dot{q}(t) + g(q(t)) = 0,
\label{eqn:swingdyn}
\end{equation*}
where, dropping time dependencies, $M(q) \in \mathbb{R}^{2 \times 2}$ is the
inertia matrix, $C(q, \dot{q}) \in \mathbb{R}^{2 \times 2}$ is the Coriolis
matrix, and $g(q) \in \mathbb{R}^2$ is the gravitational vector: 
\begin{gather*}
M(q) = \left[ \begin{smallmatrix}
		\kappa_1 + \kappa_2 + 2 \delta c_2 & \kappa_1 + \delta c_2 \\
		\kappa_1 + \delta c_2 & \kappa_1
	\end{smallmatrix} \right], 
C(q,\dot{q}) = \left[ \begin{smallmatrix}
		-\delta s_2 \dot{q}_2 & -\delta s_2(\dot{q}_1 + \dot{q}_2) \\
		\delta s_2 \dot{q}_1 & 0
	\end{smallmatrix} \right], 
g(q) = \gamma \left[ \begin{smallmatrix} \beta s_1  + s_{12} \\ s_{12}
\end{smallmatrix} \right]\!.
\end{gather*}
The notation $c_{12}$ represents $\cos(q_1 + q_2)$, $s_{12}$ is $\sin(q_1 +
q_2)$, etc.  Furthermore, the equations have been nondimensionalized such that
units of time, length, and torque are scaled by $d_t = \sqrt{\frac{\ell_1 +
\ell_2}{g}}$, $d_{\ell} = \ell_1 + \ell_2$, and $d_u = g m_2 r_2$, respectively.
These scalings define the following unitless parameters: $\beta = \frac{m_1 r_1
+ m_2 \ell_1}{m_2 r_2}$, $\delta = \frac{\ell_1}{ d_{\ell}}$, $\gamma = \frac{g
d_t^2}{d_{\ell}}$, $\kappa_1 = \frac{1}{d_{\ell}} \left(\frac{I_2}{m_2 r_2} +
r_2\right)$, $\kappa_2 = \frac{1}{d_{\ell}} \left(\frac{I_1 + m_1 r_1^2}{m_2
r_2} + \frac{\ell_1^2}{r_2}\right)$, and $\tau = \frac{t}{d_t}$.

\subsection{The Jump Map}
\label{ssec:jump}
When the robot's end effector (e.g., its hand or foot) collides with a surface,
the robot undergoes an instantaneous plastic impact.  As described in
\cite{Rosa2012}, for $\ell_1 = \ell_2$, this results in a jump map $h(x)$ such
that
\begin{equation}
h(x) = \begin{bmatrix}
A & 0 \\ 
0 & \Delta(q)
\end{bmatrix} x + \begin{bmatrix} b \\ 0 \\ 0 \end{bmatrix}, \quad
\Delta(q) = \frac{1}{a_2} \left[ 
	\begin{smallmatrix}
		a_1 - 2 \delta (\delta - \beta \kappa_1) c_2 & a_1 \\
		-4 \delta \cos(\frac{q_{2}}{2})^2 a_3 & -a_1 + 2 \delta (\kappa_2 - \beta \delta) c_2
	\end{smallmatrix}
\right],
\label{eqn:h}
\end{equation}
where $A = \left[ \begin{smallmatrix} 1 & 1 \\  0 & -1 \end{smallmatrix}
\right]$ and $b = \left[ \begin{smallmatrix} \pi \\ 0 \end{smallmatrix} \right]$
achieves a relabeling of the coordinate system; $\Delta(x) \in \mathbb{R}^2$ is
our impact map relating the pre-impact velocity $\dot{q}^-$ to the post-impact
velocity $\dot{q}^+$ such that $\dot{q}^+ = \Delta(q^-)\dot{q}^-$; and $a_1 = 2
\beta \delta \kappa_1 - 2 \kappa_1 \kappa_2 $, $ a_2 = \delta^2 - 2 \kappa_1
\kappa_2 + \delta^2 \cos(2 q_{2}) $, and $ a_3 = (\beta \kappa_1 - \kappa_2 +
(\beta - 1) \delta c_2) $.

\subsection{The Switching Function}
\label{ssec:switch}
We define the switching function $\phi$ as
\begin{equation}
\phi(x(t, x_0), t) = \begin{cases}
    \sin(\sigma(x_0) - \sigma(x(t, x_0))) & \text{for SBS} \\
    t - \tau & \text{for TBS},
\end{cases}
\end{equation}
where $\sigma(x(t)) = \frac{\pi - q_2(t)}{2} - q_1(t)$ is the instantaneous
slope at time $t$.  We expand on our definition of $\sigma$ in the sequel.  The
$\sin$ term is more straightforward to explain: slopes that differ in multiples
of $\pi$ represent the same inclined surface (e.g., level ground locomotion can
be represented with $\sigma$ values of $k\pi$, $k \in \mathbb{Z}$).

\subsubsection{Defining the Walking Surface}
As is common in the literature, a pre-defined value of $\sigma$ determines the surface incline.  In our work, we introduce the notion of an
instantaneous slope $\sigma(x) = \frac{\pi - q_2}{2} - q_1$.  This equation for
$\sigma$ measures the angle between the straight line connecting the feet (or
hands) of the robot and the horizontal.  The quantity always coincides with the
value of any pre-defined surface incline at the beginning and end of a step;
e.g., for a post-impact state $x_0$, any pre-defined slope must equal
$\sigma(x_0)$.

\begin{remark}
Typically, $q_1$ and $q_2$ are chosen so that the robot is on a pre-defined
slope $\sigma$.  Here, we simply do the converse.  Relative to other works, we
eliminate $\sigma$ as an unnecessary free parameter, especially when computing
gait families with numerical continuation.  In the end, there is no loss of
generality.  For any work that treats $\sigma$ as a free parameter, a little bit
of geometry will show that the value of $\sigma$ can be determined by the
configuration of the paper's two-link model.
\end{remark}

\section{Methodology and Results}
\label{sec:results}
The physical parameters of our study are chosen to match those of
\cite{Gomes2005}.  The approximate values are $\beta = 18.13$, $\delta_1 =
0.50$, $\delta_2 = 0.50$, $\gamma = 1$, $\kappa_1 = 0.28$, and $\kappa_2 =
8.84$.  We compute the initial gaits of each gait family using the methods
outlined in \cite{Gomes2005} for brachiating and \cite{Rosa2014b} for walking.  We then use a pseudo-arclength continuation method with a fixed step
size of 0.001.  The algorithm runs until it either fails or generates 50,000 points.
For period-one fixed points, stability is determined by assessing the magnitude
of the eigenvalues of the linearized impulsive dynamics.  To study the long-term
evolution of unstable gaits with respect to switching strategy, we simulate a
gait across 1,000 impacts and then analyze the impact sequence.  The code used to
generate these solutions and visualize the bifurcations is available at
\href{https://github.com/aestr6/TWOLINK\_NODYCON}{https://github.com/aestr6/TWOLINK\_NODYCON}.

\subsection{The Big Picture}

\begin{figure}[t]
\centering
\includegraphics[width=\linewidth]{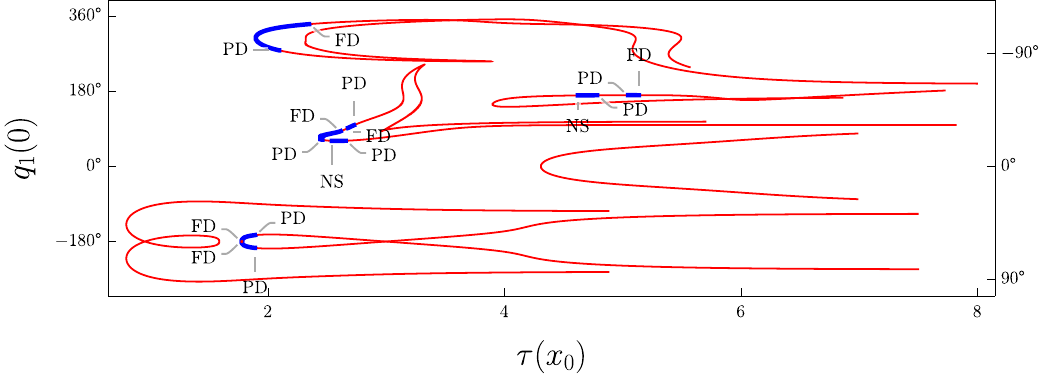}
\hfill
\includegraphics[width=\linewidth]{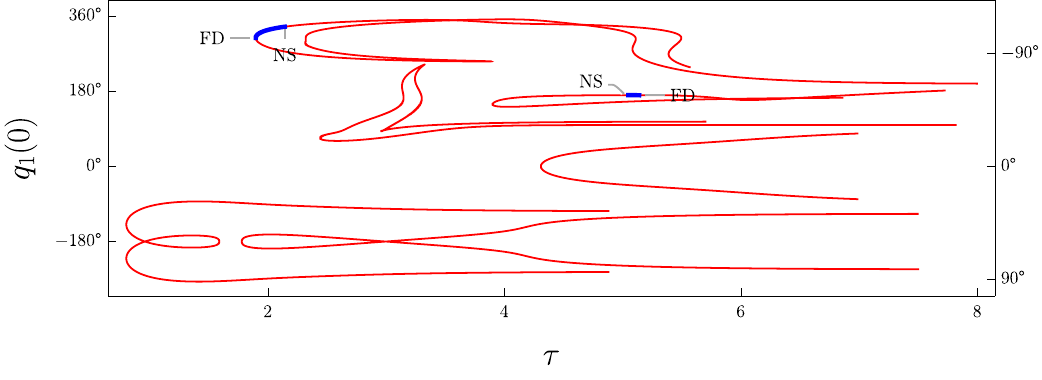}
\caption{(top) Example gait families with their stability properties (red
unstable, blue stable) and bifurcation points under SBS.
(below) A similar plot, but with gaits subject to TBS.  To
avoid overlapping curves, we plot joint angles beyond a range of $2 \pi$ on the
left axes of each plot.  The corresponding values for $q_1(0) \in (-\pi, \pi]$ under modulo arithmetic
are shown on the right axes.}

\label{fig:famstab}
\end{figure}

Figure~\ref{fig:famstab} provides a global view of passive dynamic families of
walking and brachiating gaits as curves projected onto a
switching-time-joint-angle plane.  To distinguish between TBS and SBS, we use
$\tau$ to denote an independent switching time parameter and $\tau(x_0)$ as a
state-dependent switching time, respectively.

Overall, there are four types of qualitatively similar motions that can be found
within each gait family. Each motion corresponds to walking or brachiating
above or below a surface (e.g., the example gaits of Figure~\ref{fig:motions}).
A more nuanced exploration of gait families with respect to these motions can be
found in \cite{Rosa2013b}.  With respect to switching strategies, the plots
highlight differences in gait stability and bifurcations.  For example, three of
the curves have regions of stable gaits under an SBS strategy and contain the
three simple bifurcation types.  In contrast, TBS only has two gait families
with stable regions.  These regions are smaller and undergo either
NS or FD bifurcations once the gaits are no longer
stable.  In the remaining sections, we further explore the top-most curve and
the bottom right-most curve in these plots.

\subsection{Exploring a Brachiating Family of Gaits}

\begin{figure}[t]
    \centering
    \begin{tikzpicture}

        \node[anchor=center] (basicPDB) at (-4.3, -0.05) {\includegraphics[scale=0.22]{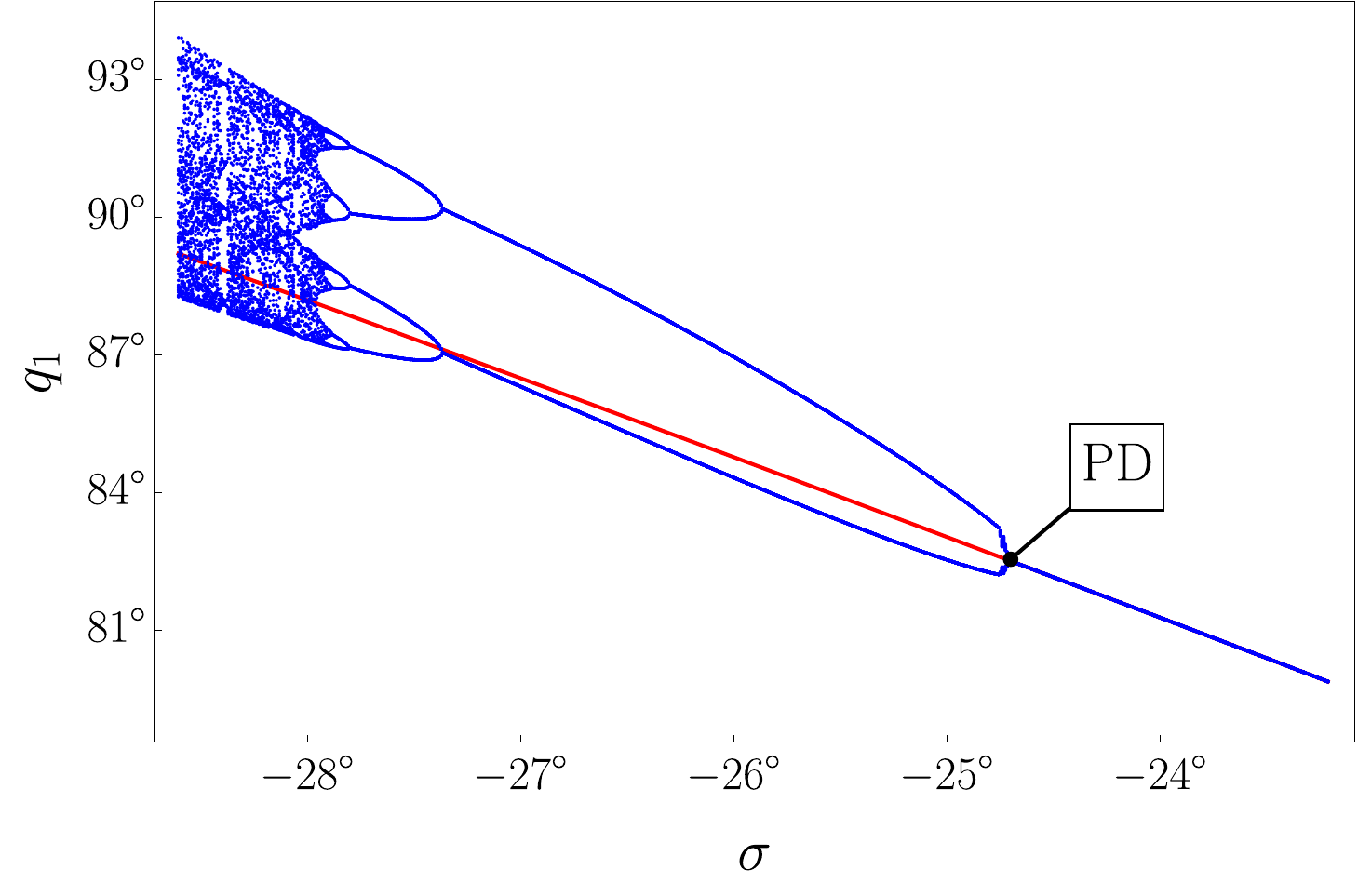}};
        \node[anchor=center] (mainBase) at (1.95,0.15) {\includegraphics[width=0.53\textwidth]{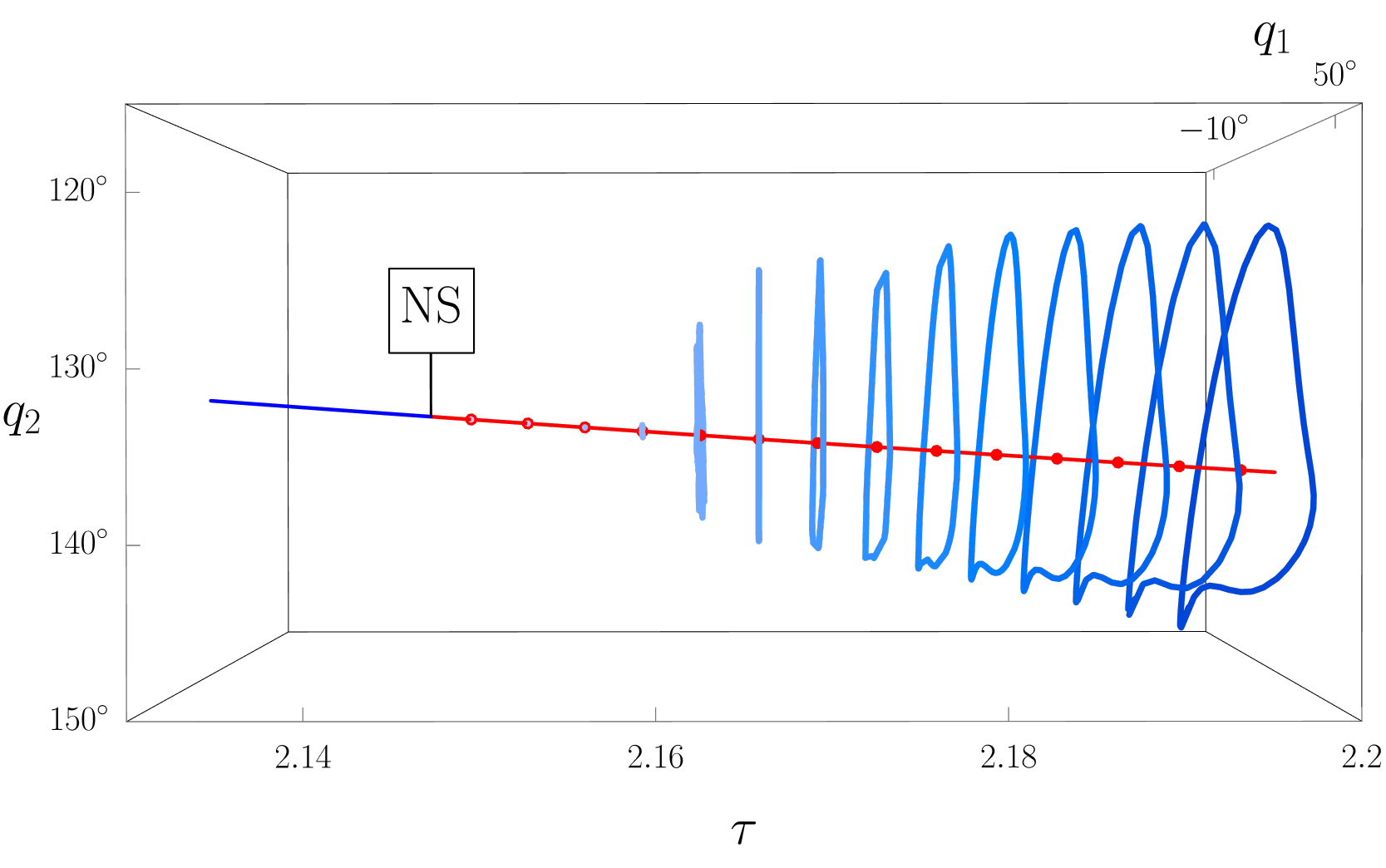}};
        
        \node[anchor=center] (neimarkSlice) at (2.7, -3.8) {\includegraphics[scale=0.2]{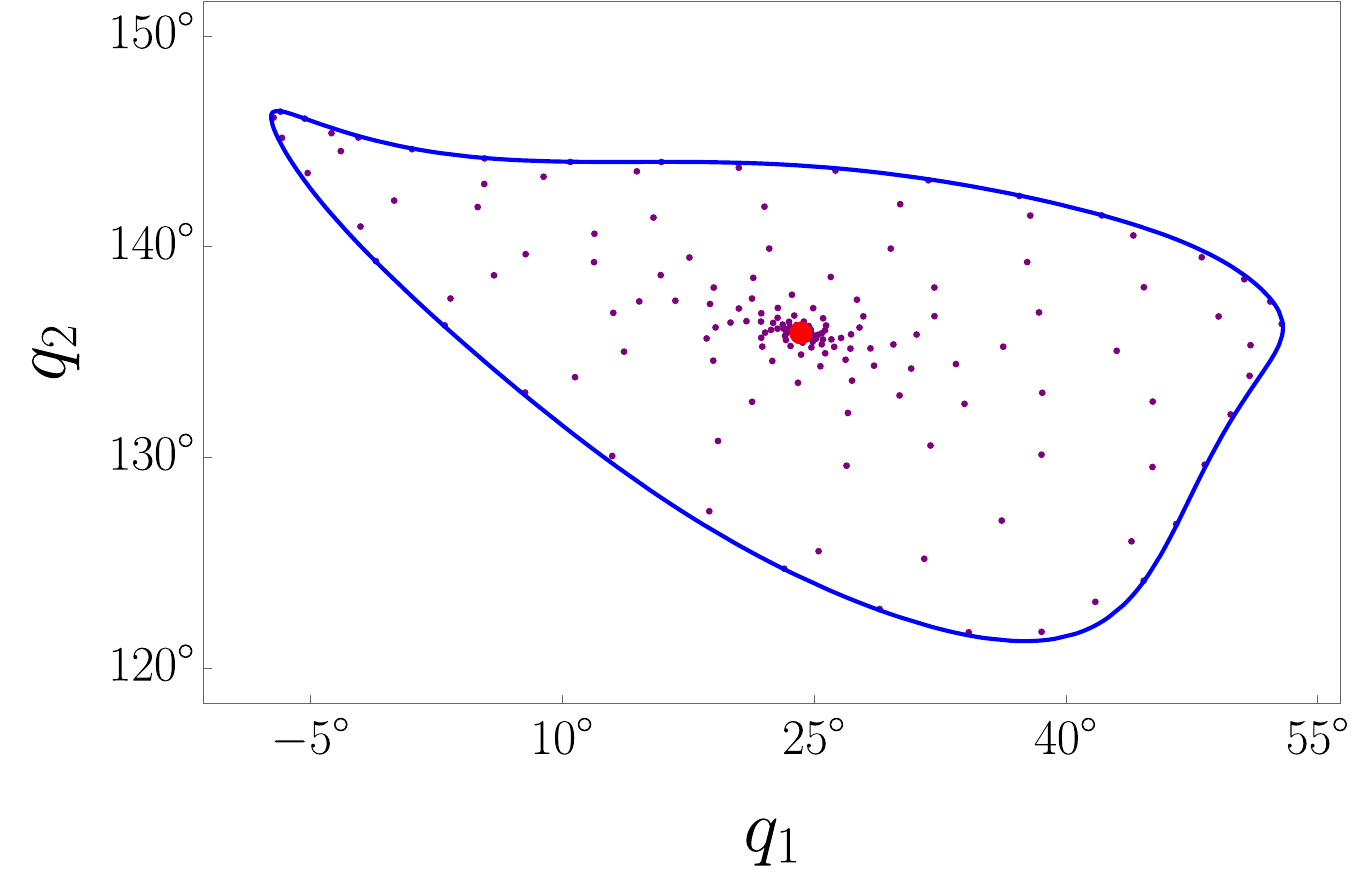}};
        \node[anchor=center] (neimarkEig) at (-1.5, -3.8) {\includegraphics[scale=0.2]{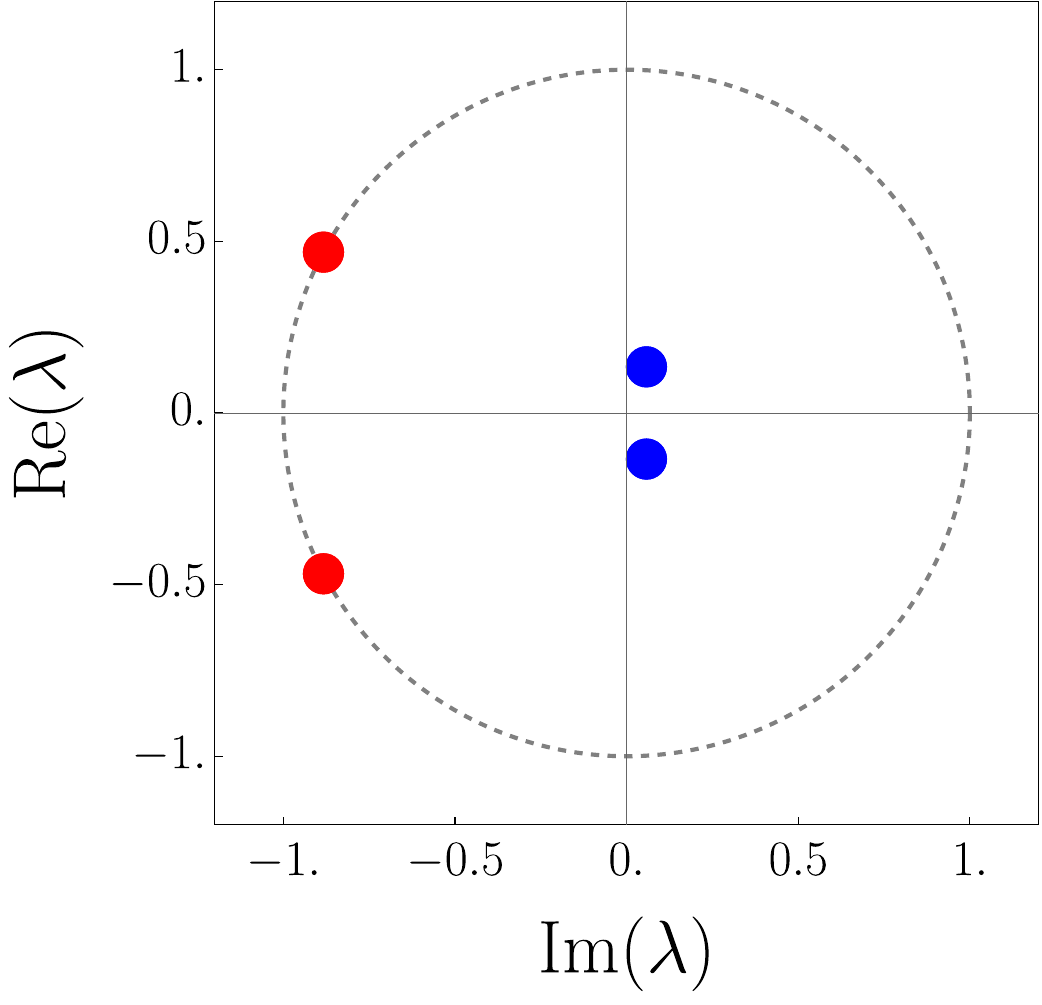}};
        \node[anchor=center] (pdbEig) at (-5.3, -3.8) {\includegraphics[scale=0.2]{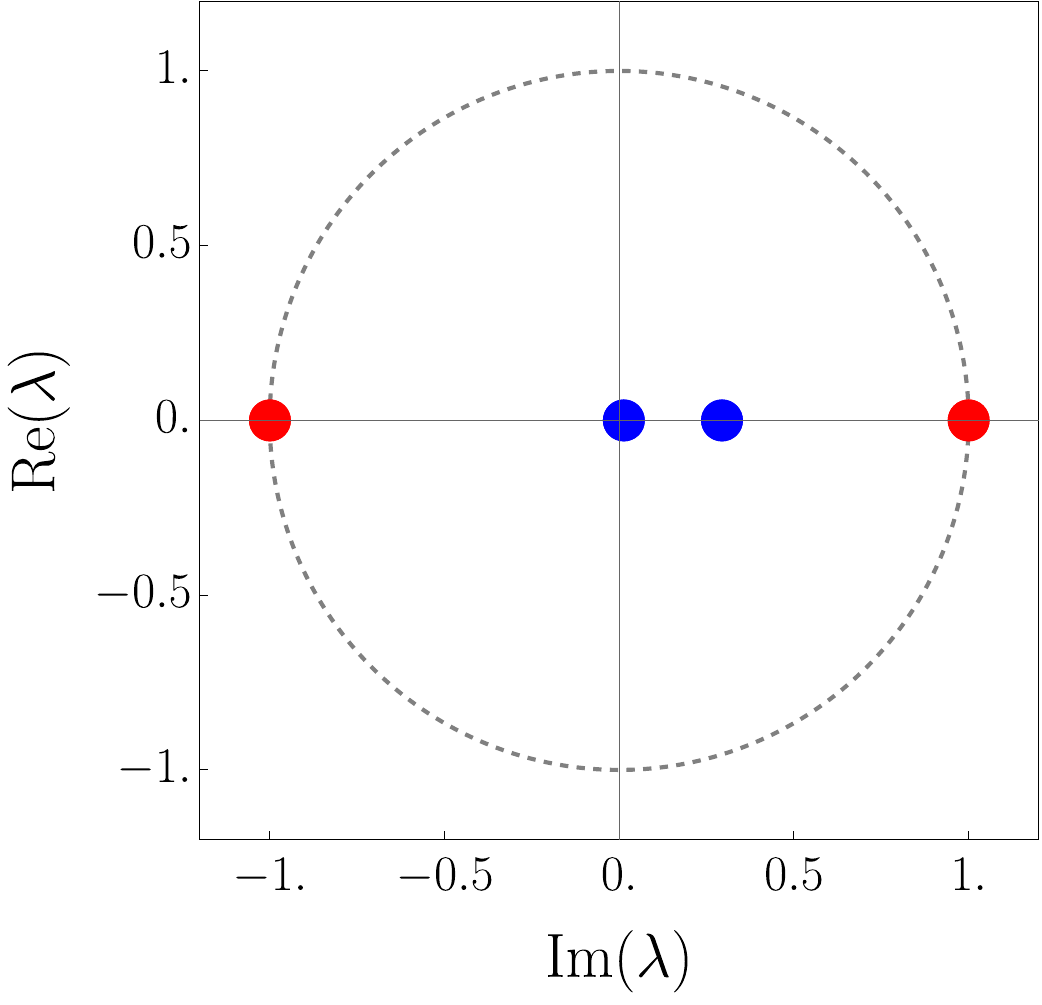}};

        \draw[->, thick] (-4.97,-2.1) to[bend left=25] (-3.0,-0.49);
        \draw[->, thick] (-1.15,-2.1) -- (0.725,0.23);
        \draw[->, thick] (3.15,-3.4) to[bend right=15] (3.645,0.01);
    \end{tikzpicture}
    \caption{(top left) A brachiating branch of gaits (part of the top-most
curve in Figure~\ref{fig:famstab}) undergoing a PD bifurcation
under an SBS strategy. (top right) The same branch undergoing an NS bifurcation under a TBS strategy. (bottom left) The
corresponding eigenvalues are displayed for each bifurcation.  (bottom right) A state-space
projection of the evolution of an unstable period-one gait (red) converging to a
closed invariant curve (blue) during the NS bifurcation, with 1,000 impacts
plotted (purple).}

    \label{fig:bifurcations}
\end{figure}

The top-most curve in Figure~\ref{fig:famstab} is a curve of brachiating gaits
with gaits along the curve that will qualitatively look like the motions in the
bottom row of Figure~\ref{fig:motions}.  Figure~\ref{fig:bifurcations} compares the evolution of gaits on the curve under SBS as slope is varied and under TBS as switching time is varied.  For slopes ranging from $-2.0^\circ$ to
$-24.5^\circ$, under SBS, the period-one fixed points are stable, resulting in
gaits that remain on the initial gait family.  At
$\sigma_\text{crit}\approx-24.57^\circ$, the stable period-one gait loses
stability through a PD bifurcation, giving rise to a stable
period-two gait. This transition is confirmed by an eigenvalue crossing the unit
circle at $-1$. Note that SBS always has a marginally stable eigenvalue at $+1$,
as expected for hybrid systems with state-defined switching surfaces
\cite{Bainov1993}.

The same brachiating family exhibits different stability properties under TBS.
From $\tau = 1.90$ to $2.44$, the period-one fixed points are stable along the
brachiating branch.  At $\tau_\text{crit} \approx 2.14$, an NS bifurcation
occurs. The bifurcation is confirmed by the complex conjugate eigenvalues
crossing the unit circle, with values $\lambda_{1,2} \approx -0.88\pm 0.47i$.
Beyond the NS bifurcation, trajectories converge onto closed invariant curves.
The bottom-right plot in Figure~\ref{fig:bifurcations} shows 1,000 successive
impacts spiraling away from the unstable fixed point toward a stable closed
invariant curve.

\subsection{Fold Bifurcations and Symmetry in Time-Based Switching}
\begin{figure}[t]
    \centering
    \includegraphics[width=\linewidth]{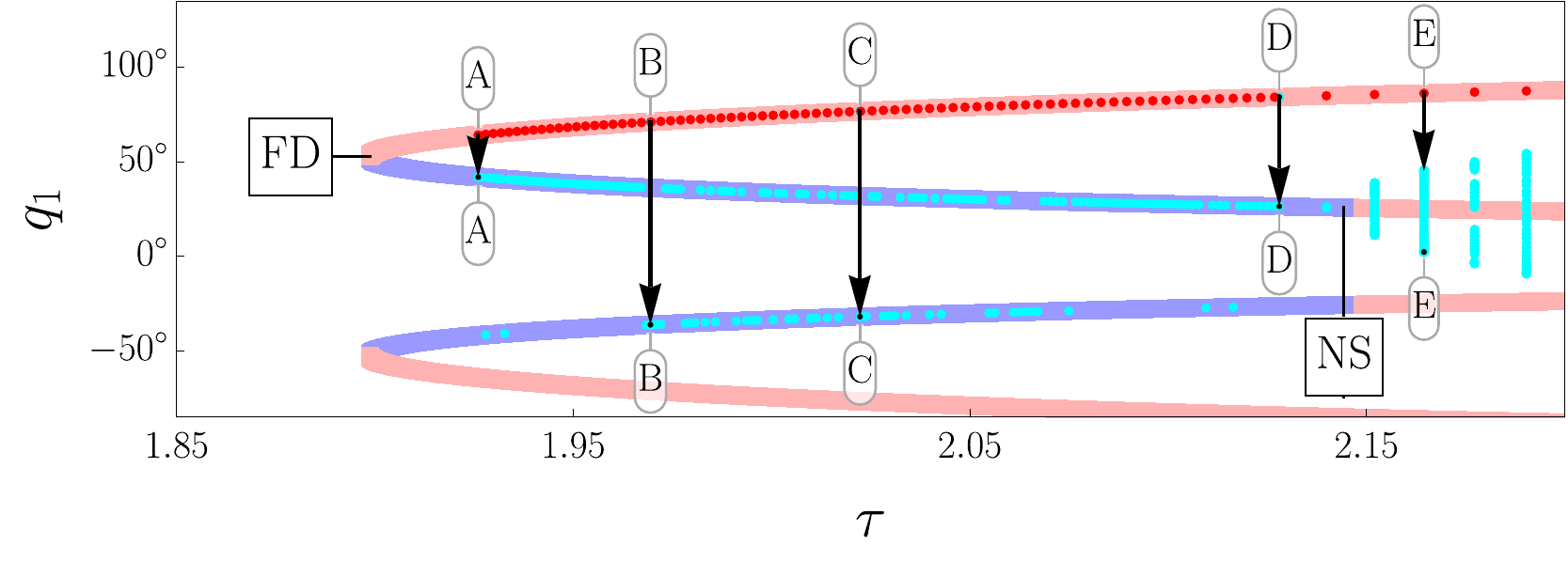}
    \caption{Convergence of unstable points after an FD bifurcation for
the brachiating branch under a TBS method.}
    \label{fig:symmetry}
\end{figure}

The NS bifurcation is only one boundary of the stable region shown in Figure
\ref{fig:bifurcations}. The opposite end exhibits an FD bifurcation, shown in
Figure~\ref{fig:symmetry}. This bifurcation occurs at $\tau_\text{crit} \approx
1.92$ and is characterized by the crossing of the unit circle by an eigenvalue
at +1.  Figure~\ref{fig:symmetry} depicts the evolution of post-impact states
following this bifurcation.

Due to the bilateral symmetry of the system, if $(x_0, \tau, \sigma)$ is a gait,
then so is $(-x_0, \tau, -\sigma)$.  The resulting motion is identical, but
simply reflected (e.g., a gait that walks to the left instead walks to the
right).  Consequently, every gait family naturally possesses a mirrored
counterpart.  This can be seen in Figure~\ref{fig:symmetry}, which shows a
segment of a brachiating curve and its mirrored segment below the $q_1 = 0$
axis.  Notably, trajectories originating near the unstable regions of one family
are shown to converge into either the stable domain of their original or mirrored
gait families along directions of constant $\tau$. For instance, at label A
($\tau=1.93$) and D ($\tau=2.13$), the gaits converge onto the stable region of
the primary curve.  Alternatively, at B ($\tau=1.97$) and C ($\tau=2.02$), the
gaits converge onto the stable region of the mirrored curve. This behavior
highlights the complex basins of attraction that arise from the
system's geometric symmetries.

Beyond the stable region, no stable period-one solutions remain. Consequently,
the trajectories converge to closed invariant curves near the NS bifurcation
point.

\subsection{Exploring a Walking Family of Gaits After a Fall}

\begin{figure}[t]
\centering
\includegraphics[width=0.95\linewidth]{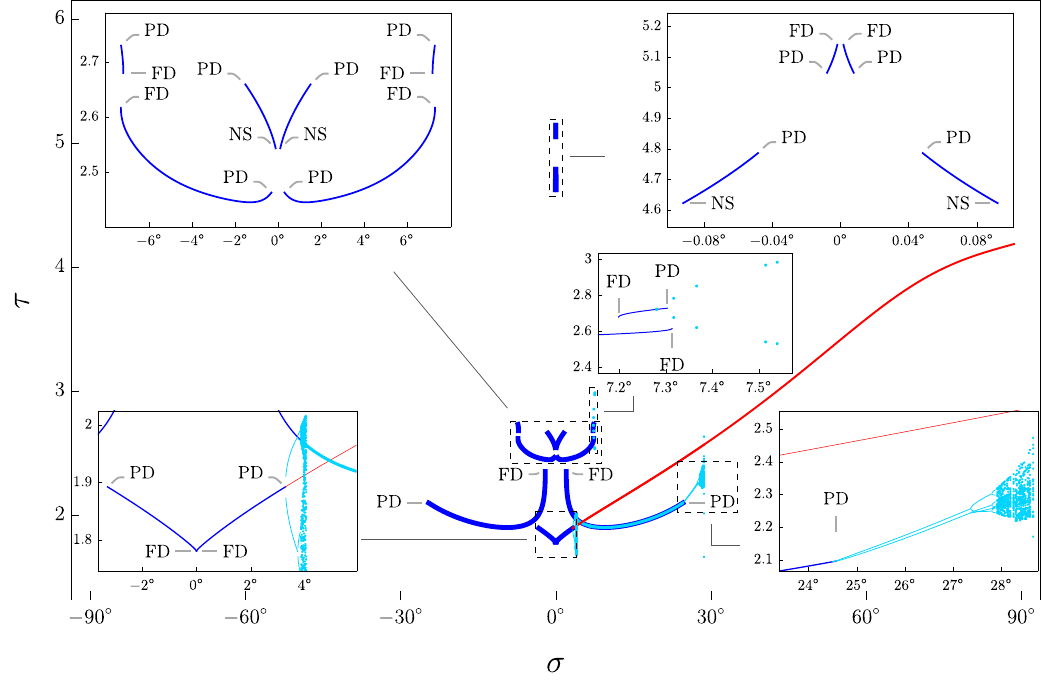}
\caption{The evolution of unstable period-one walking gaits (UWGs) along a
segment in red after 1,000 impacts (points in cyan) under SBS
(gaits evolve along lines of constant slope in the plot).  Stable period-one
walking and brachiating gaits are in blue.  The points in cyan are also stable
periodic orbits.  The insets show zoomed in regions of interest.  For example,
the inset of the bottom-left corner shows UWGs converging to stable walking
gaits of period-$2^k$ ($k > 1 \in \mathbb{Z}$), and the inset in the
bottom-right corner shows UWGs converging to stable period-$2^k$ brachiating
gaits.  Most UWGs converge to stable period-one brachiating gaits.}

\label{fig:walk}
\end{figure}

Given Figure~\ref{fig:famstab}, we see that stable walking gaits only exist for
SBS.  While walking gaits under SBS are the most explored in the literature, we
present results on an unexplored avenue of research. If we continue applying the
SBS policy, where do unstable walking gaits (UWGs) go if the robot is in
environments that allow it to transition between locomoting above and below a
surface (e.g., power lines \cite{Mbewe2025})?

Figure~\ref{fig:walk} addresses this question for the parameters of our model.
The plot shows the gait families under SBS from Figure~\ref{fig:famstab} projected onto a
slope-switching-time plane.  Instead of terminating the simulation after
the robot falls, we continue to impact the surface and plot the results after
1,000 impacts.  The final impacts are shown in cyan.  UWGs near the stable
walking region ($\sigma \in [3.29^\circ, 3.86^\circ]$) settle into a PD
periodic orbit.  After the PD structure breaks down, the UWGs generally converge
to stable period-one brachiating gaits ($\sigma \in [4.06^\circ, 24.47^\circ]$).  When the region of stable period-one brachiating gaits ends, the
UWGs then converge to the PD orbits of brachiating gaits after crossing the PD
bifurcation point on the brachiating curve at $\sigma \approx 24.57^\circ$.

While most UWGs converge to the period-one brachiating gait family below the
UWG branch, there are UWGs that converge to brachiating motions
above the unstable curve near $\tau \approx 2.7$.  The inset plot in the middle
provides a snapshot of five of these gaits.  The stable gaits are either
period-one or period-two fixed points in the region.

\section{Conclusion}
\label{sec:conc}
The comparative analysis of SBS and TBS
strategies reveals fundamental differences in the stability of passive dynamic
walking and brachiating gaits.  Most notably, we demonstrate that SBS induces
FD, PD, and NS bifurcations, whereas TBS
results in NS and FD bifurcations.  Additional results include the discovery of intertwined basins of attraction for mirrored sets of gaits for models with bilateral symmetry under TBS. Furthermore, we identify basins of attraction that exist for unstable period-one walking gaits under SBS after the robot falls, which converge to stable brachiating gaits.

To the best of our knowledge, this is the first combined analysis showing the
connection between walking and brachiating gaits in the space of passive dynamic
walking gaits.  However, we stopped short of investigating the chaotic regimes
present with our model.  Future work will be dedicated to formally analyzing
routes to chaos and identifying potential strange attractors in the gait space.

\bibliographystyle{bibtex/spmpsci_unsrt}
\bibliography{nodycon2026}

\end{document}